\documentclass[10pt]{article}
\usepackage{latexsym}
\newtheorem{theorem}{Theorem}
\newtheorem{lemma}{Lemma}

\newcounter{llista}
\newcounter{llista1}

\renewcommand{\arraystretch}{1.5}
\def\be{\begin{equation}}
\def\ee{\end{equation}}
\def\bea{\begin{eqnarray}}
\def\eea{\end{eqnarray}}

\begin{document}

\author{B. Coll \thanks{ 
       D\'epartement d'Astronomie Fondamentale UMR-8630 CNRS,
       Observatoire de Paris, \, 61 Av. Observatoire; F-75014 Paris, France},
        J. Llosa\thanks{Departament de F\'{\i}sica Fonamental, Universitat de 
      Barcelona, Diagonal, 647, E-08028 Barcelona, Spain; e-mail address: 
      pitu@ffn.ub.es} and D. Soler}
      
\title{Three-Dimensional Metrics as Deformations\\ of a Constant Curvature
Metric}  
\maketitle

\begin{abstract}
Any three-dimensional metric \,$g$\, may be locally obtained from
a constant curvature metric, \,$h$\,, by a deformation like
$$ g = \sigma h + \epsilon \,s \otimes s \: \: , $$
where \,$\sigma$\, and \,$s$\, are respectively  a  scalar and a
one-form, the sign \,$ \epsilon = \pm 1$\, and a functional relation
between \,$\sigma$\, and the Riemannian norm of \,$s$\, can be arbitrarily
prescribed.  The general interest of this result in geometry and
physics, and the related open problems, are stressed.

\end{abstract}

\section{Introduction}
 It is known, since an old result by Riemann \cite{Riem},  that a
$n$-dimensional metric  has \,$f = n(n-1)/2$\, {\em degrees of freedom},
that is, it is locally equivalent to the giving of \,$f$\, functions.  As
this feature is related to some particular choices of local charts, which
are obviously non-geometric objects \cite{Geom}, it seems to be
generically a not covariant property.

According to it, a two-dimensional metric  has \,$f = 1$\, degrees of
freedom. In this case, however, a stronger result holds, as it is well
known \cite{Conf}, namely: 
{\it any two-dimensional metric \,$g$\, is locally conformally flat, 
\,$g = \phi\,\eta  $\,, \,$\phi$\, being the
{\rm conformal} deformation factor and \,$\eta$\,the flat metric.}

Contrarily to what the above Riemann's general result suggests, the
two-dimensional case is {\em intrinsic} and {\em covariant}, i.e. it only
needs the knowledge of the  metric \,$g$\, and only involves tensor
quantities, specifically, the sole degree of freedom is represented by a
{\em scalar},  the conformal deformation factor \,$\phi$\,.

The question thus arises of, whether or not, for \,$n>2$\, there exist
similar intrinsic and covariant local relations between an arbitrary
metric \,$g$\,, on the one hand, and the corresponding flat one
\,$\eta $\, together with a set of  \,$f$\, covariant quantities on the
other. 

To our knowledge, no result of this type has been published. Indeed, the
known results concerning the diagonalization of any three-dimensional
metric \cite{Diag},\cite{UnivDef} do not belong to this type. As a matter
of fact, besides the \,$f = 3$\, scalars and the (more o less implicit) flat
metric, these results {\em also} involve a particular ortogonal triad of
vector fields.
Also, in the context of the General Theory of Relativity, such
a $n$-dimensional relation has  been proposed by one of us, but
unfortunately it remains for the moment only  a mere conjecture.

In this paper we shall answer affirmatively the
three-dimensional case. This dimension is the solution to the equation
\,$f = n$\,, so that one is tempted to take (the components of) a vector
field as the covariant set (of \,$f=3$\,  quantities). On the other hand,
the result being deliberately local, it would seem that the essentials of
the flat metric in this matter is its minimal freedom, i.e. the maximal
dimension of its isometry group, so that it should be possible to
substitute it by  a prescribed constant curvature metric. We shall see
that both assumptions work.

In fact, the paper is devoted to prove the following main result:

\begin{theorem}
Any three-dimensional Riemannian metric \,$g$\, may be locally obtained from
a constant curvature metric \,$h$\, by a  deformation of the form
\be  \label{e0}
 g = \sigma h + \epsilon\,\mbox{\boldmath$s$}\otimes\mbox{\boldmath$s$} \: \: , 
\ee
where $\sigma$\, and \,$\mbox{\boldmath$s$}$\, are respectively a scalar function
and a differential 1-form; the sign $ \epsilon = \pm 1$ and a relationship
$\Psi(\sigma, |s|)$\, between the scalar \,$\sigma$\, and the Riemannian
norm \,$|s|$\, may be arbitrarily prescribed. 
\end{theorem}

This result should be interesting in geometrical as well as in
physical situations.

In geometry, perhaps one of the first questions to be answered is
the following: In two dimensions it is known that the gauge of
the conformal factor \,$\sigma$\, or, equivalently, the set of flat
metric tensors conformal to a given metric is given by
the solutions of the Laplacian, \, $\Delta\sigma = 0$\,
\cite{Lapla}. In the three-dimensional case here considered,
what is the gauge of the  vector fields \,$ s$\, associated to
a given metric \,$g$\,? or, equivalenty, how many constant
curvature metric tensors \,$h$\, correspond to \,$g$\, through the
relation (\ref{e0})?

But many other interesting questions arise. For instance,
since the theorem states a correspondence between a metric
\,$g$\, and a couple $\,(\sigma,s$, what conditions must $\,\sigma\,$ and
$\,s\,$ fulfill in order that \,$g$\, admits a continuous group of
symmetries? 

In classical physics, the above theorem should be useful in
(finite) deformation theory of materials; equation (\ref{e0}) may
be considered as an {\em ideal universal deformation law},
allowing, from an unconstrained or not initial state (described in
material coordinates by the tensor \,$h$\,),  to reach any other
deformation state (described in the same coordinates by the tensor
\,$g$\,) \cite{LeyPlana}. This ideal universal law allows to
associate, to every deformation state of a material, a vector
field \,$s$\, among those of the  gauge class of the flat
metric.

In general relativity, any vacuum space-time is locally equivalent
to its Cauchy data, \,$\{ g,K\}$\,, \,$ g$\, being the spatial
metric and \,$K$\, the extrinsic curvature of the initial instant.
These data have to verify the {\em constraint equations}, a set of
four equations for which  many years ago Lichnerowicz showed
\cite{Lich} that to every arbitrarily given metric \,$\tilde g$\,
it corresponds a unique solution \,$\{ g,K\}$\, such that \,$g =
\sigma \tilde g$\,. This beautiful result is however useless 
for precise physical situations because, \,$ g$\, being
initialy  unknown, one does not see how to choose the good
starting metric \,$\tilde g$\,, which has to give \,$ g$\, by
conformity. Such an objection may be eliminated using  (\ref{e0}) in
the constraint equations. Our theorem also allows to translate
notions such as asymptotic flatness or spatial singularity in
terms of the differential 1-form \,$ s$\, over a flat metric \,$h$\,.

The paper is organized as follows. Sections 2 and 3 are devoted
to proof the above theorem  and some examples of this result
are presented in section 4.

\section{Flat deformation of a given metric \label{s2}}
Instead of proving theorem 1 as stated in the introduction, we shall prove
the following equivalent result:

\begin{theorem}
Let $({\cal V}, g)$ be a Riemannian 3-manifold. There locally exist a
function $\phi$ and a differential 1-form $\mu$ such that the tensor 
\be
\tilde g :=\phi\,g -\epsilon\,\mbox{\boldmath$\mu$}\otimes\mbox{\boldmath$\mu$}
\label{e1}
\ee
(with $\epsilon =\pm 1$) is also a Riemannian metric with constant
curvature. Besides, an arbitrary relation between $\phi$ and
$|\mu|^2:=g^{ij}\mu_i\mu_j$ can be imposed in advance. 
\end{theorem}

The equivalence between both theorems follows immediately on substituting 
$$ h = \tilde{g} \,, \qquad \sigma = \phi^{-1} \,, \qquad
\mbox{\boldmath$s$} = \phi^{-1/2} \mbox{\boldmath$\mu$} $$
into equation (\ref{e0}). The present formulation (\ref{e1}) stresses that
we seek to derive \,$\tilde g$\, from a given \,$g$.

The proof spreads over sections \ref{s2} and \ref{s3} and is based on the
comparison of the Riemannian geometries respectively defined by $g$ and
$\tilde g$. 

We start by considering the Riemannian connexions $\nabla$ and $\tilde\nabla$.
In an arbitrary frame $\{e_i\}_{i=1,2,3}$ the expression (\ref{e1}) reads:
\be
\tilde g_{ij}:=\phi\,g_{ij} - M_{ij} \;, \qquad {\rm with} \qquad
M_{ij} :=\epsilon\,\mu_i\,\mu_j
\label{e3}
\ee
We shall consider the difference tensor:
\be
B^j_{ki} :=\tilde\gamma^j_{ki} - \gamma^j_{ki}
\label{e4}
\ee
which is symmetric:
\be
B^j_{ki} = B^j_{ik}
\label{e2}
\ee
because both connexions are torsion free.

Now, since $\nabla_k g_{ij} = \tilde\nabla_k \tilde g_{ij} =0\,$ and
taking (\ref{e2}) into account, we easily obtain that:
\be
B^j_{ik} = \frac12\,\left[\phi_k\,g_{ir} + \phi_i\,g_{kr} - \phi_r\,g_{ik} -
      \nabla_k\,M_{ir} -\nabla_i\,M_{kr} +\nabla_r\,M_{ik} \right]
      \,\tilde h^{rj}
\label{e6}
\ee
where
\be
\tilde h^{rj} := \phi^{-1} \,\left( g^{rj} +\frac{1}{\phi-m_0}M^{rj} \right)
\;, \qquad {\rm with} \qquad m_0:= g^{ij} M_{ij} = \epsilon|\mu|^2 \,,
\label{e7}
\ee
is the inverse metric for $\tilde g_{ij}$. (Indices are always raised,
resp., lowered, with the metric $g^{ij}$, resp., $g_{ij}$, and the
notation $\tilde g^{rj}$ is reserved to $\tilde g_{il} g^{ir} g^{lj}$.)

\subsection{The curvature tensors \label{ss21}}
The curvature tensor for $\tilde g$ is \cite{Choq}
\be
\tilde R^j_{\;\;ikl} = e_k\tilde\gamma^j_{li} - e_l\tilde\gamma^j_{ki} +
\tilde\gamma^j_{km} \tilde\gamma^m_{li} - \tilde\gamma^j_{lm} \tilde\gamma^m_{ki} -
c^m_{kl} \tilde\gamma^j_{mi}
\label{e8}
\ee
For a 3-manifold this tensor is equivalent to the tensor (the one can be
obtained from
the other)
\bea
\tilde G^{ij} & = &\frac14\, \tilde\eta^{ikl}\,\tilde\eta^{jsr}\,\tilde R_{srkl} \nonumber   \\
 & = & \frac12\,\tilde\eta^{ikl}\,\tilde\eta^{j\;\;r}_{\;\;s}\,\left(e_k\tilde\gamma^s_{lr}+
\tilde\gamma^s_{km} \tilde\gamma^m_{lr} - \tilde\gamma^m_{kl} \tilde\gamma^s_{mr} \right)
\label{e9}
\eea
where $\tilde\eta^{ikl}$ is the contravariant volume tensor associated to
$\tilde g$. This $\tilde G^{ij}$ is related to the Einstein tensor. Indeed, in
three dimensions \cite{EISSEN2}:
$$\tilde R_{srkl} = \tilde g_{rk} \tilde R_{sl} +\tilde g_{sl}\tilde R_{rk}
- \tilde g_{rl} \tilde R_{ij} - \tilde g_{sk} \tilde R_{rl} +
\frac{\tilde R}{2}\,(\tilde g_{sk}\tilde g_{rl} - \tilde g_{sl}\tilde g_{rk})
$$
whence it follows immediately that: $\tilde G^{ij} = \tilde R^{ij} -
\frac12 \, \tilde h^{ij} \,\tilde R\,$.

Similar relations hold for the curvature tensor
$R^j_{\;\;ikl}$, the metric $g_{ij}$ and the volume tensor $\eta^{ikl}$.

Using equations (\ref{e9}) and (\ref{e4}), and the fact that the volume
tensors $\tilde\eta^{ikl}$ and $\eta^{ikl}$ are proportional ---see
appendix A--- we obtain:
\be
\tilde G^{ij} = D^2 G^{ij} +\frac{D^2}{2}\,\eta^{ikl}\,\eta^{j\;\;r}_{\;\;s}\,
\left(\nabla_k B^s_{lr}+ B^s_{km} B^m_{lr} \right)
\label{e10}
\ee
where the relationships (\ref{A2a}) and (\ref{A3}) in Appendix A:
$$ \tilde\eta^{ikl} = D\,\eta^{ikl} \;, \qquad  \qquad
D^2:=\frac{\det g}{\det\tilde g} = \frac{\phi^{-2}}{\phi - m_0}  $$
have been taken into account.

The condition that $\tilde g$ has constant curvature \cite{EISSEN3} is:
$$ \tilde R_{jikl} = K \,(\tilde g_{jk}\tilde g_{il} - \tilde g_{jl} \tilde g_{ik}) \; ,
\qquad \qquad K = {\rm constant} \, ,$$
which in terms of $\tilde G^{ij}$ reads:
\be
\tilde H^{ij}:= \tilde G^{ij} - K\,\tilde h^{ij} = 0
\label{e10a}
\ee

\subsection{The second Bianchi identity \label{ss22}}
In terms of the tensor $\tilde G^{ij}$, the second Bianchi identity
\cite{EISSEN4} reads
\be
\tilde\nabla_i\tilde G^{ij} \equiv 0
\label{e12}
\ee
which, using the tensor $\tilde H^{ij}$ introduced above and taking into
account that $K$ is a constant, leads to:
$$ \tilde\nabla_i\tilde H^{ij} \equiv 0  \,;$$
In terms of the connexion $\nabla$ and the difference tensor $B^j_{ik}$, this identity can be also written as:
\be
\nabla_i\tilde H^{ij} + B^i_{il}\tilde H^{lj} + B^j_{il}\tilde H^{li} \equiv 0
\label{e13}
\ee
and must be understood as follows:

\begin{lemma}
Let $g\,$, $\phi\,$ and $M_{ij}\,$ be, respectively, a metric, a scalar
function and a symmetric tensor such that $\tilde g$ defined by (\ref{e3})
is regular. Then the
tensor field $\tilde G^{ij}(g,\gamma,\phi,M)$ defined by (\ref{e10})
satisfies identically (\ref{e13}).
\end{lemma}

In the next section we shall consider the condition (\ref{e10a}) as a partial
differential system on the unknowns $\mbox{\boldmath$\mu$}_i$ and $\phi$. The problem of
solving this system is pretty similar to solve
Einstein equations in 3 dimensions.

\section{The Cauchy problem \label{s3}}
Let ${\cal S}_0$\, be a surface in a 1-parameter family of surfaces ${\cal
S}_\lambda \subset {\cal V}_3$, let $n$ be the unit $g$-normal vector and
let $\{e_i\}_{i=1,2,3}$ be a $g$-orthonormal tetrad adapted to ${\cal
S}_\lambda$, i. e., $e_3=n$\,.

We have to solve the second order partial differential system (\ref{e10a})
with $\tilde G^{ij}$ given by (\ref{e10}). We first notice that the set of
three equations:
\be
\tilde H^{3j}:= \tilde G^{3j} - K\,\tilde h^{3j} = 0 \;, \qquad j=1,2,3
\label{e14}
\ee
do not contain second order normal derivatives of the unknowns:
neither $\nabla_3\nabla_3M_{kl}$ nor $\nabla_3\nabla_3\phi$. Indeed,
$$\eta^{3kl}\nabla_k=\sum_{b=1}^2\eta^{3bl}\nabla_b $$
only involves tangential derivatives, i. e., along $e_1$ and $e_2$. (Hereon
the indices $a,\,b,\,c,\,\ldots$ run from 1 to 2, whereas the indices
$i,\,j,\, k,\,\ldots$ run from 1 to 3.)

On the other hand, the remaining three equations:
\be
\tilde H^{ab}:= \tilde G^{ab} - K\,\tilde h^{ab} = 0 \;, \qquad a,b=1,2
\label{e15}
\ee
do contain second order normal derivatives. After a short calculation,
taking (\ref{e3}), (\ref{e6}), (\ref{e7}) and (\ref{e10}) into account,
we readily obtain that eq. (\ref{e15}) can be written as:
\be
\tilde H^{ab}:= \frac{D^2}{4} \,\epsilon^{a3l}\,\epsilon_{bsr}\,\tilde{h}^j_s
\,\left( \delta^3_r\,\delta^i_j-\delta^3_j\,\delta^i_r \right)\,\left(
\ddot{\phi} \, \delta_{il} - \epsilon\,\mbox{\boldmath$\mu$}_i\,
\ddot{\mbox{\boldmath$\mu$}}_i -
\epsilon\,\mbox{\boldmath$\mu$}_l\,\ddot{\mbox{\boldmath$\mu$}}_i
\right) + P^{ab} = 0
\label{e16}
\ee
where a dot means the covariant normal derivative $\nabla_3$, and $P^{ab}$
does not depend on second order normal derivatives.

We have however three equations and four unknowns, hence the problem is, at
this stage, underdetermined.
We can thus introduce an arbitrary additional relation:
\be
\Psi(\phi,m_0) = 0
\label{e17}
\ee
which will be hereafter referred to as {\it gauge}.

By successive differentiation along $\nabla_3$, this constraint induces
other differential constraints, namely,
\bea
\nabla_3\Psi & := & \Psi_1\,\dot\phi + \Psi_2\,2\epsilon\,
\mbox{\boldmath$\mu$}^i\dot{\mbox{\boldmath$\mu$}_i} = 0
\label{e17a} \\
\nabla_3\nabla_3\Psi & := & \Psi_1\,\ddot\phi + \Psi_2\,2\epsilon\,
\mbox{\boldmath$\mu$}^i\ddot{\mbox{\boldmath$\mu$}}_i + P_0 = 0
\label{e17b}
\eea
where:
$$ \Psi_1 :=\frac{\partial\Psi}{\partial\phi} \qquad {\rm and} \qquad
\Psi_2:= \frac{\partial\Psi}{\partial m_0}  \,. $$
and $P_0$ does not depend on second order normal derivatives.

Now, the second order partial differential system
\be
\left. \begin{array}{l}
       \tilde H^{ab} = 0 \;, \qquad  a,b=1,2 \\
       \nabla_3\nabla_3\Psi = 0
       \end{array}
       \right\}
\label{e18}
\ee
is quasilinear and has four equations for four unknowns.

Its characteristic determinant is:
$$
\Delta_1 = \left|\begin{array}{cccc}
    \frac{-D^2(2\phi-m_0-\epsilon(\mbox{\boldmath$\mu$}_1)^2)}{4\phi(\phi-m_0)} &
    \frac{D^2\mbox{\boldmath$\mu$}_1\,(\mbox{\boldmath$\mu$}_2)^2}{4\phi(\phi-m_0)} &
    \frac{D^2\epsilon \mbox{\boldmath$\mu$}_2(4\phi-2 m_0-\epsilon(\mbox{\boldmath$\mu$}_2)^2)}{4\phi(\phi-m_0)} &
    0 \\
    \frac{D^2\mbox{\boldmath$\mu$}_1\mbox{\boldmath$\mu$}_2}{4\phi(\phi-m_0)} &
    -\frac{D^2\epsilon \mbox{\boldmath$\mu$}_2(2\phi- m_0-\epsilon(\mbox{\boldmath$\mu$}_2)^2)}{4\phi(\phi-m_0)} &
    -\frac{D^2\epsilon \mbox{\boldmath$\mu$}_1(2\phi- m_0+\epsilon(\mbox{\boldmath$\mu$}_2)^2)}{4\phi(\phi-m_0)} &
    0 \\
   -\frac{D^2(2\phi-m_0-\epsilon(\mbox{\boldmath$\mu$}_2)^2)}{4\phi(\phi-m_0)} &
   \frac{D^2\epsilon \mbox{\boldmath$\mu$}_1(4\phi-2 m_0-\epsilon(\mbox{\boldmath$\mu$}_2)^2)}{4\phi(\phi-m_0)} &
   \frac{D^2\mbox{\boldmath$\mu$}_2\,(\mbox{\boldmath$\mu$}_1)^2}{4\phi(\phi-m_0)} &
   0 \\
   \Psi_I &  2\epsilon \Psi_2 \mbox{\boldmath$\mu$}_1 &
   2\epsilon \Psi_2 \mbox{\boldmath$\mu$}_2  & 2\epsilon \Psi_2 \mbox{\boldmath$\mu$}_3
               \end{array}  \right|
$$
or
\bea
\Delta_1 & =& \frac{D^6}{2^4}\mbox{\boldmath$\mu$}_3\,\Psi_2 \epsilon
\left[(\mbox{\boldmath$\mu$}_1)^2+(\mbox{\boldmath$\mu$}_2)^2\right] \,\left[2\phi - 2 m_0 +
\epsilon(\mbox{\boldmath$\mu$}_3)^2  \right] \nonumber \\
 & & \left[2\phi - m_0 - \epsilon(\mbox{\boldmath$\mu$}_2)^2 \right]
\,\left[2\phi -  m_0\right]
\label{e19}
\eea
\renewcommand{\arraystretch}{1.1}

${\cal S}_0$ is non-characteristic \cite{JOHN71} if $\Delta_1\neq 0$ that,
taking (\ref{A4}) into account (see the Appendix) ---which ensures that both $g$ and
$\tilde{g}$ are nondegenerate and positive---, reduces to:
\be
\Psi_2 = \frac{\partial\Psi}{\partial m_0} \neq 0\;, \qquad \mbox{\boldmath$\mu$}_3 \neq 0
 \quad {\rm and} \quad \mbox{\boldmath$\mu$}_1^2+\mbox{\boldmath$\mu$}_2^2 \neq 0
\label{e20}
\ee

As a consequence, we have shown the following result
\begin{theorem}
{Let  ${\cal S}_0 \subset {\cal V}$ be a surface  and $\{e_i\}_{i=1,2,3}$
a $g$-orthonormal frame adapted to ${\cal S}_0$, and let us be given:
\begin{list}
{(\alph{llista})}{\usecounter{llista}}
\item {a gauge $\Psi(\phi,m_0)$ and}
\item {a set of Cauchy data:
    $$\overline{\mbox{\boldmath$\mu$}_i}=\left.\mbox{\boldmath$\mu$}_i\right|_{{\cal S}_0} \;, \quad
      \overline{\phi}=\left.\phi\right|_{{\cal S}_0} \;, \quad
      \overline{\dot{\mbox{\boldmath$\mu$}}_i}=
        \left.\nabla_3\mbox{\boldmath$\mu$}_i\right|_{{\cal S}_0} \;, \quad
      \overline{\dot\phi}=\left.\nabla_3\phi  \right|_{{\cal S}_0} $$
    (where a bar means ``the value on ${\cal S}_0$) such that:
  \begin{list}
  {{\bf (S\arabic{llista1})}}{\usecounter{llista1}}
  \item \hspace*{2em} $\overline{\mbox{\boldmath$\mu$}_3} \neq 0 \;,  \quad
        \mbox{\boldmath$\mu$}_1^2+\mbox{\boldmath$\mu$}_2^2 \neq 0 \;, 
        \quad \overline{\phi} > 0 \;, \quad
        {\rm and} \quad \overline{\phi} - \overline{m_0} > 0 \;,$
  \item the gauge does depend on $m_0$, that is,
        $\displaystyle{\left.\frac{\partial\Psi}{\partial m_0}
        \right|_{{\cal S}_0}  \neq 0} $
  \item equations (\ref{e17a}) and (\ref{e17b}) hold on ${\cal S}_0$, and
  \item the subsidiary conditions:
  $\left.\tilde H^{3j}\right|_{{\cal S}_0} = 0
  \;$, $j=1,2,3$
  \end{list}}
\end{list}
We can then find a solution $\mbox{\boldmath$\mu$}_l$, $\phi$ defined on a neighbourhood
${\cal U}$ of ${\cal S}_0$ such that fulfills (\ref{e10a}) and
(\ref{e17}), i. e.:
$$ \tilde{H}^{ij} (\phi, \mbox{\boldmath$\mu$}_l) = 0\,, \quad i,j=1,2,3 \qquad {\rm and}
  \qquad \Psi(\phi, m_0) = 0 \;.$$}
\end{theorem}

\paragraph{Proof:}
Indeed, by conditions {\bf S1} and {\bf S2} above, ${\cal S}_0$ is
non-characteristic, the Cauchy-Kovalevski theorem \cite{JOHN71}
can be applied and a solution $\phi$, $\mbox{\boldmath$\mu$}_l$ of the partial differential
system (\ref{e18}) can be found in a neigbourhood ${\cal U}_1$ of ${\cal
S}_0$ fulfilling conditions {\bf S1} through {\bf S4}.

Now, let $\tilde g_{ij}$ be the metric constructed on
${\cal U}_1$ by substitution of the solution $\phi$ and $\mbox{\boldmath$\mu$}_i$ into (\ref{e3}).
Let us see that $\tilde g_{ij}$ has constant curvature, i. e., $\tilde
H^{ij} = 0\,$, for $i,j =1,2,3$ on a neigbourhood of ${\cal S}_0$.

We have on the one hand that $\tilde H^{ab} = 0\,$, for $a,b =1,2$, because
$\phi$ and $\mbox{\boldmath$\mu$}_i$ is a solution of (\ref{e18}). And, on the other, the
remaining three equations, namely, $\tilde H^{3j} = 0\,$,  $j =1,2,3$,
hold on ${\cal S}_0$ (condition {\bf S4}).

To prove that the latter condition propagates well to a neigbourhood of ${\cal
S}_0$, we separate the normal and tangential derivatives in the Bianchi
identity (\ref{e13}) and, taking into account that (\ref{e18}) holds on
${\cal U}_1$, we arrive at:
\be
\nabla_3\tilde H^{3j} + \sum_{b=1}^2 \left[e_b\tilde H^{b3}\,\delta_3^j +
 2\,B^j_{3b} \tilde H^{3b} + \Gamma^b_{b3} \tilde H^{3j}+
 \Gamma^j_{b3} \tilde H^{b3} \right] +
  B^i_{i3}\tilde H^{3j} + B^j_{33} \tilde H^{33}  =  0
\label{e23}
\ee
The latter can be taken as a linear homogeneous partial differential system for
the unknown $\tilde H^{3j}$ which, for the Cauchy data expressed by
condition {\bf S4} has the unique solution $\tilde H^{3j} = 0\,$ on a
neigbourhood ${\cal U}_2$ of ${\cal S}_0$. Hence,
$$\tilde H^{ij} = 0\;, \qquad {\rm on} \qquad {\cal U} ={\cal U}_1 \cap
{\cal U}_2 $$
and $\tilde g_{ij}$ has constant curvature in ${\cal U}$.

It is obvious that the gauge condition $\Psi(\phi,m_0) =0$ also propagates
to the neigbourhood of ${\cal S}_0$, as a consequence of the last equation in
(\ref{e18}) and the conditions {\bf S3}.

\subsection{The subsidiary conditions \label{ss31}}
We shall now see whether the subsidiary conditions (\ref{e14}) are not too
restrictive. In the adapted $g$-orthonormal frame $\{e_i\}_{i=1,2,3}$
introduced at the begining of this section, the conditions (\ref{e14}) and
the second of equations (\ref{e17a}) read:
\be
{\eta^{3ab}} \nabla_a({B^s_{br}}{\eta^{j\;\,r}_{\;\,s}}) +
{\eta^{3ab}}\,{\eta^{j\;\,r}_{\;\,s}}\,{B^s_{am}}
{B^{m}_{br}} + 2\,{G^{3j}} -
\frac{2K}{{D}^2} \,{\tilde h^{3j}} \approx 0
\label{e24}
\ee
\be
{\Psi}_I \, {\dot\phi} + 2\,{\Psi_2}\,{\mbox{\boldmath$\mu$}^i}\,
{\dot{\mbox{\boldmath$\mu$}}_i} \approx 0
\label{e25}
\ee
where ``$\approx$" means that the equality holds on ${\cal S}_0$.

The latter equations yield four relations to be fulfilled by the Cauchy data
and can be used as a partial differential system on ${\cal S}_0$ to
determine part of the Cauchy data, namely: $\overline{\dot\phi}\,$ and
$\overline{\dot{\mbox{\boldmath$\mu$}}_i}\,$, in terms of $\overline{\phi}\,$ and $\overline{\mbox{\boldmath$\mu$}_i}\,$.

Making explicit the terms containing $\nabla_a \overline{\dot\phi}\,$ and
$\nabla_a \overline{\dot{\mbox{\boldmath$\mu$}}_i}\,$, equations (\ref{e24}) and (\ref{e25})
respectively yield:
\be
A^{ja} \,\nabla_a\overline{\dot\phi} + A^{lja}\,\overline{\dot{\mbox{\boldmath$\mu$}}_l} + Z^j \approx 0
\label{e26}
\ee
and
\be
2\overline{\mbox{\boldmath$\mu$}^i}\,{\Psi_2}\,\nabla_a\overline{\dot{\mbox{\boldmath$\mu$}}_i} +
{\Psi_I}\,\nabla_a\overline{\dot\phi} + Z \approx 0
\label{e28}
\ee
where $Z$ and $Z_j$ depend only on $\overline{\phi}$ and 
$\overline{\mbox{\boldmath$\mu$}_i}$, their
derivatives tangential to ${\cal S}_0$ and on $\overline{\dot\phi}$ and
$\overline{\dot{\mbox{\boldmath$\mu$}}_i}$, but not on tangential derivatives of the latter.

Furthermore, a short calculation yields:
\bea
A^{ja}&=&\frac1{2{\overline\phi}}\left[\left(2+
\frac{\epsilon({\overline{\mbox{\boldmath$\mu$}_3}})^2}{\overline\phi -
{\overline{m}_0}}\right)\,\delta^{ja} +
\frac{\epsilon}{\overline\phi -{\overline{m}_0}}\,{\overline{q}^a}\,{\overline{q}^j} -
\frac{\epsilon\,{\overline{\mbox{\boldmath$\mu$}_3}}}{\overline\phi
-{\overline{m}_0}}\,{\overline{\mbox{\boldmath$\mu$}^a}}\,\delta_3^j \right]
\label{e27a} \\
  &  &  \nonumber \\
A^{lja} & = &\frac{\epsilon}{2{\overline\phi}}\left[ 2\,
\overline{\mbox{\boldmath$\mu$}^a}\,
{\eta^{lj3}} + \frac{\epsilon\,\overline{\mbox{\boldmath$\mu$}_3}}{\overline\phi -
{\overline{m}_0}}\,
\overline{\mbox{\boldmath$\mu$}^a}\,{\eta^{ljs}}\,
\overline{\mbox{\boldmath$\mu$}_s} - 
{\eta^{3al}}\,{\overline{q}^j}\,\frac{2\overline\phi
-{\overline{m}_0}}{\overline\phi -{\overline{m}_0}}\, \right. - \nonumber \\
      & & \left.  \frac{\epsilon}{\overline\phi
-{\overline{m}_0}}\,{\overline{q}^a}\,{\overline{q}^j}\,
\overline{\mbox{\boldmath$\mu$}_l}\right]
\label{e27b}
\eea
where $\overline{q}^i:=\eta^{3ij}\overline{\mbox{\boldmath$\mu$}_j}$.

To put the Cauchy problem for the partial differential system
(\ref{e26}-\ref{e28}) on ${\cal S}_0\,$, let ${\cal C}\subset{\cal S}_0\,$
be a given curve and $\tau = \tau^a\,e_a \in T{\cal S}_0$, the unit vector
orthogonal to ${\cal C}$. Assume that an adapted frame is chosen in
$T{\cal S}_0$, such that
$\tau^1=0$ and $\tau^2=1$, then the curve is non-characterisitc if, and
only if, the characteristic determinant does not vanish, i. e.:
\be
\Delta_2:= - \frac{\epsilon\,
\overline{\mbox{\boldmath$\mu$}_2}\,\overline{\mbox{\boldmath$\mu$}_3}^2\,\overline{\mbox{\boldmath$\mu$}_1}^2\,{\Psi_2}\,
[2\overline\phi- 2{\overline{m}_0} +\epsilon\,\overline{\mbox{\boldmath$\mu$}_3}^2]\,
(2\overline\phi -{\overline{m}_0})}{2\,\overline\phi^3\,(\overline\phi -{\overline{m}_0})^3} \neq 0
\label{e29}
\ee
That is, if, and only if, the data $\overline{\phi}$, $\overline{\mbox{\boldmath$\mu$}_i}$ on
${\cal S}_0$ are given such that:
\begin{list}
{{\bf (S\arabic{llista})}}{\usecounter{llista}}
\item \hspace*{2em} $\Psi(\overline{\phi},\overline{m}_0) = 0 \quad {\rm
   and} \quad  \overline{\mbox{\boldmath$\mu$}_i} \neq 0 \;, \quad i=1,2,3\;$
\item \hspace*{2em} and \,$\displaystyle{\frac{\partial \Psi}{\partial 
\overline{m}_0}
\neq 0\quad}$ on ${\cal S}_0$.
\end{list}

\subsection{Summary}
What we have proved so far is that for any given:
\begin{list}
{(\alph{llista})}{\usecounter{llista}}
\item surface ${\cal S}_0$ and curve ${\cal C} \in {\cal S}_0$,
\item gauge function $\Psi(\phi,m_0)$ and
\item {data: $\overline{\phi}\,, \overline{\mbox{\boldmath$\mu$}_i}\,, i=1,2,3$ on
      ${\cal S}_0$ and $\overline{\dot\phi}\,, 
      \overline{\dot{\mbox{\boldmath$\mu$}}_i}\,,
      i=1,2,3$ on ${\cal C}$, such that
      $$ \Psi_2(\overline{\phi},\overline{m}_0) \neq 0 \,, \quad
       \overline{\phi} >0 \,, \quad \overline{\phi} -\overline{m}_0 >0 \;
       \mbox{ and } \overline{\mbox{\boldmath$\mu$}_j} \neq 0 \,, j=1,2,3 $$}
\end{list}
the quasilinear partial differential system (\ref{e26}) and (\ref{e28})
can be integrated to determine  $\overline{\dot\phi}\,$ and
$\overline{\dot{\mbox{\boldmath$\mu$}}_j}\,, j=1,2,3$ on a surface ${\cal S}_1$ (a
neighbourhood of ${\cal C}$ on ${\cal S}_0$).

Then the data  $\overline{\phi}\,$, $\overline{\mbox{\boldmath$\mu$}_i}\,$,
$\overline{\dot\phi}\,$ and $\overline{\dot{\mbox{\boldmath$\mu$}}_j}\,$, $i,j=1,2,3$, on the
surface ${\cal S}_1$ fulfill the conditions {\bf S1} through {\bf S4} of
theorem {\bf 2}. Hence, functions $\phi$ and $\mbox{\boldmath$\mu$}_i$ on a 3-dimensional
neighbourhood ${\cal U}$ of ${\cal S}_1$ can be obtained such that
(\ref{e10a}) and (\ref{e17}) are fulfilled.

Therefore the metric $\tilde{g}$ obtained by substituting these $\phi$
and $\mbox{\boldmath$\mu$}_i$ in equation (\ref{e1}) has constant curvature.  \hfill $\Box$

\section{Two examples \label{S4}}
For the sake of illustration we shall consider two cases of 3-dimensional
Riemannian manifolds and locally deform them into flat metrics, in the
sense stated in Theorem 2. Since both cases exhibit some symmetries, the
solutions will be rather proposed than derived by solving the partial
differential system presented in sections 2 and 3.

\subsection{Schwarzschild space \label{SS4.1}}
The title is a shortening for {\it the space 3-manifold for Schwarzschild
coordinates in Schwarzschild spacetime}. The metric is:
\be
\hat{g} = \kappa^{-1}\,dr\otimes dr + r^2\,d\theta\otimes d\theta + 
r^2\,\sin^2\theta\,d\varphi\otimes d\varphi
\label{e4.1}
\ee
with $\kappa = 1 - \frac{2m}r $, in the region $r>2m$ (otherwise the
metric is not Riemannian).

This metric can be deformed into a flat metric in several ways. Among others:
\begin{description}
\item[\ref{SS4.1}.A]{Choosing $s = \sqrt{\kappa^{-1} - 1}\,dr$, we readily obtain:
         $$ \hat{g} = \tilde{g} + s\otimes s $$
         where $\tilde{g}=dr\otimes dr + r^2\,d\theta\otimes d\theta + 
         r^2\,\sin^2\theta\,d\varphi\otimes d\varphi$ is flat.}
\item[\ref{SS4.1}.B]{It is well known tha,t changing $r$ into the coordinate
         $$ R = \frac12\left(r\,\sqrt{\kappa} + r - m\right)
         \qquad r = R \left(1+\frac{m}{2R}\right)^2 \,,$$
         the metric becomes:
         $\,\hat{g} = \sigma \,\tilde{g}\,$,
        where 
        $$\sigma = \left(1+\frac{m}{2R}\right)^4 \qquad{\rm and}\qquad
        \tilde{g} := dR\otimes dR + R^2\,d\theta\otimes d\theta +
         R^2\,\sin^2\theta\,d\varphi\otimes d\varphi $$ 
         is a flat metric.}
\end{description}

\subsection{Kerr space \label{SS4.2}}
In Boyer-Lindquist coordinates, the Kerr metric is \cite{MTW}:
$$g =  -\frac{\Delta}{p^2}\,\left[dt - a\,\sin^2\theta\,d\varphi \right]^2
      + \frac{\sin^2\theta}{p^2}\,\left[a\,dt - (r^2+a^2)\,d\varphi \right]^2 +
      \frac{p^2}{\Delta}\,dr^2 + p^2\,d\theta^2
$$
where $\,p^2=r^2 + a^2\cos^2\theta\,$ and $\,\Delta=r^2-2mr+a^2\,$.

The stationary space 3-manifold is endowed with the Riemannian metric:
$$
\hat{g} = \frac{p^2}{\Delta}\,dr\otimes dr + p^2\,d\theta\otimes d\theta +
\frac{p^2\Delta}{p^2-2mr}\,\sin^2\theta\,d\varphi\otimes d\varphi
$$
or, equivalently,
\be
\hat{g} = \frac{p^2}{r^2}\,\left(\frac1{\Delta}\,dr\otimes dr 
+ r^2\,d\theta\otimes d\theta +
\frac{\Delta}{p^2-2mr}\,r^2\sin^2\theta\,d\varphi\otimes d\varphi \right)
\label{e4.3}
\ee
in the region $r^2-2mr+a^2\cos^2\theta>0\,$ (otherwise the
metric is not Riemannian).
Similarly as in the case {\bf B} above, this last expression suggests to
define:
$$ R = \frac12\,\left(\sqrt{\Delta} + r - m \right)  \qquad
   r = R \left[\left(1+\frac{m}{2R}\right)^2 -\frac{a^2}{4R^2}\right]  $$
Then (\ref{e4.3} can be  written as:
\be
\hat{g} = \frac{p^2}{R^2}\,\left(\tilde{g} +
\frac{a^2\,R^2\sin^4\theta}{p^2-2mr}\,d\varphi^2 \right)
\label{e4.4}
\ee
where $\tilde{g} = dR\otimes dR + R^2\,d\theta\otimes d\theta + 
     R^2\sin^2\theta\,d\varphi\otimes d\varphi $,
which has already the form (\ref{e0}) with $\epsilon=+1$ and
$$ \sigma =\frac{p^2}{R^2} \qquad{\rm and} \qquad 
   s = \frac{a\,p\,\sin^2\theta}{\sqrt{p^2-2mr}}\,d\varphi $$

\section{Conclusion and outlook \label{S5}}
We have shown that, locally, any Riemannian 3-dimensional metric $g$ can
be deformed along a direction $s$ into a metric $\sigma h$ that is
conformal to a metric of constant curvature, as stated in theorem {\bf 1}.

The direction $s$ is not uniquely determined by the metric $g$, and the
decomposition (1) can be achieved in an infinite number of ways.
Determining more precisely the class of $\sigma$ and $s$ which deform a
given $g$ into a constant curvature metric $h$ will be the object of
future work. Specially the case where both, $g$ and $h$, are flat.

It is tempting to look at (\ref{e0}) as an equivalence relation, namely,
\begin{quotation}
$g_1 \sim g_2$ if, and only if, $\exists \sigma$ and $s$ such that:
$g_1=\sigma\,g_2 + \epsilon\,s\otimes s \,$.
\end{quotation}
However, this relation is not in general transitive.

Expression (\ref{e0}) can also be taken as the definition of a
transformation of a metric $h$ into a metric $g$, somewhat reminding of
Kerr-Schild transformations \cite{KSHM}. Namely, given $\sigma
\in\Lambda^0({\cal V})$, $\epsilon =\pm 1$ and
$\,s\in\Lambda^1({\cal V})\,$ we can define the transformation:
$$ T_{\sigma,\epsilon,s}: g_1 \longmapsto g_2 = T_{\sigma,\epsilon,s}[g_1]
= \sigma\,g_1 + \epsilon \, s\otimes s \,.$$ 
which acts on Riemannian metrics on the 3-manifold ${\cal V}$. It is
obvious that the identity and the inverse of $T$ belong to this class of
transformations, but the product $T_{\sigma,\epsilon,s} \circ
T_{\sigma^\prime,\epsilon^\prime,s^\prime}$ do not in general belong to this
class. One could ask however whether this class contains some groups of
transformations.

Finally, it would be interesting to extend theorem 1, or a similar result,
to a higher number of dimensions. In this sense it seems worth to pursue the
conjecture of {\it universal deformation law} \cite{UnivDef}, namely, any
Riemannian (resp., semi-Riemannian) metric $g$ can be written as:
$$ g_{\alpha\beta} = \sigma \, h_{\alpha\beta} + \mu
F_{\alpha\lambda}F^\lambda_\beta $$
where $F_{\alpha\beta}$ is a 2-form, $h$ is a flat metric and $\sigma$ and
$\mu$ are scalar functions of $F$.
(Actually, expression (\ref{e0}) is a particular case of the above
conjecture for $n=3$, with $F_{ij} = \eta_{ijk}\,s^k$,
for $i,j,k=1,2,3\,$.

\section*{Acknowledgement}
The work of J. Ll. and of D. S. is supported by DIGICyT, contract 
no. BFM2000-0604 and by Institut d'Estudis Catalans (S.C.F.).

\section*{Appendix A: The volume tensors}
Let us choose  be a $\tilde g$-orthonormal frame, $\{\tilde
e_i\}_{i=1,2,3}$, such that $M_{ij} = A\,\delta^3_i\,\delta^3_j$. 
From (\ref{e1}) we have that: 
\be
g_{ij}= \phi^{-1} \,\left(\delta_{ij} + A\,\delta^3_i\,\delta^3_j\right)  
\;, \qquad\qquad \tilde g_{ij} = \delta_{ij}
\label{A1}
\ee
whence it follows that
\be
D^2:=\frac{\det g}{\det\tilde g} = \phi^{-3}\, (1+A)
\label{A1a}
\ee
Now, $A$ is related to the invariant $m_0:= g^{ij} M_{ij}$. Indeed,
$$ m_0 = g^{33}\,A =\frac{\phi}{1+A} \, A $$
that yields $A=m_0\,(\phi-m_0)^{-1}\,$, which substituted into (\ref{A1a})
yields:
\be
D^2 = \frac{\phi^{-2}}{\phi - m_0}
\label{A2a}
\ee

Now, let $\{\tilde \omega^j\}_{j=1,2,3}$ be the dual basis. The
contravariant volume tensors for $\tilde g$ and $g$ are, respectively,
$$
\tilde\eta^{ikl} = \epsilon^{ikl}  \qquad  {\rm and} \qquad
\eta^{ikl} = D^{-1}\,\epsilon^{ikl}
$$
where $\epsilon^{ikl}$ is the Levi-Civitt\`a symbol. From the latter it
follows immediately that:
\be
\tilde\eta^{ikl} = D\,\eta^{ikl} \;,
\label{A3}
\ee

Since both, $g$ and $\tilde g$, are Riemannian, i. e., non-degenerate and
positive, it turns out that $\phi >0$ and $D^2>0$, which, taking
(\ref{A2a}) into account, amounts to:
\be
\phi >0 \qquad {\rm and} \qquad   \phi > m_0
\label{A4}
\ee


\begin{thebibliography}{99}
\bibitem{LeyPlana} In this case, the deformation of material coordinates does
not change  the flat character of the metric, only its tensor
expression (Saint Venant equations).
\end{thebibliography}
\end{document}